\documentclass{JINST}
\pdfoutput=1

\usepackage{graphicx}
\usepackage{amssymb,amsmath}

\title{Aging studies of Micromegas prototypes for the HL-LHC.}

\author{J. Gal\'an$^a$, D. Atti\'e$^a$, J. Derr\'e$^a$, E. Ferrer-Ribas$^a$, A. Giganon$^a$, I. Giomataris$^a$, F.~Jeanneau$^a$, J.~Manjarr\'es$^a$, R. de Oliveira$^b$, P. Schune$^a$, M. Titov$^a$, J. Wotschack$^b$   \\
\llap{$^a$}CEA Saclay,\\
Gif-sur-Yvette, 91191 cedex, France \\
\llap{$^b$}CERN,\\
Geneva, Switzerland \\
  E-mail: \email{javier.galan.lacarra@cern.ch}}

\abstract{ The micromegas technology is a promising candidate to replace the forward muon chambers for the luminosity upgrade of ATLAS. The LHC accelerator luminosity will be five times the nominal one, increasing background and pile-up event probability. This requires detector performances which are currently under study in intensive R\&D activities. Aging is one of the key issues for a high-luminosity LHC application. For this reason, we study the properties of resistive micromegas detectors under intense X-ray radiation and under thermal neutrons in different CEA-Saclay facilities. This study is complementary to those already performed using fast neutrons.  }


\begin{document}

\section{Introduction}

High amplification gains are required in MicroPattern Gaseous Detectors (MPGD) in order to achieve optimum signal to noise ratios, increasing the performance of detectors in terms of energy resolution and efficiency. The gain applied allows to observe signals from gas ionizing interactions which produce few primary electrons at the detector conversion region. The populated electron avalanches achieved with these few primary electrons entail the risk to produce a discharge at the cathode of the detector when the critical electron density of $\sim 10^7 -  10^8$ electrons per avalanche is reached (Raether limit~\cite{bib7}). Once a particular detector gain has been fixed, different type of interactions inside the chamber (or produced in the chamber structure) can produce a spark with a certain probability which depends on the number of primary electrons generated~\cite{bib6}.

\vspace{0.2cm}

Discharges might affect the detector response in different ways; \emph{reducing its operating lifetime} due to intense currents produced in short periods of time, heating and melting the materials at the affected regions, \emph{damaging the read-out electronics} which have to support huge current loads in a brief period of time, and moreover \emph{increasing the detector dead-time} given that spark phenomena entail the \emph{discharge of the cathode} and therefore the amplification field is lost during a relatively long period of time, which is required by the high voltage power supply to restore the charges and recover the nominal voltage.

\vspace{0.2cm}

It was first observed in RPC-type\footnote{Resistive Plate Chambers} detectors that the introduction of a high impedance resistive coating at the anode limits the detector current during a time interval of at least some $\mu$s, constraining the spark process to the streamer phase and reducing the total amount of charge released~\cite{bib0}. Furthermore, the limited discharge current affects the field locally, remainning in the other detector regions, and thus reducing the dead-time of the detector.

\vspace{0.2cm}

Micromegas detectors were introduced in 1995~\cite{bib1} as a good candidate for high particle flux environments, and spark studies with detectors based on micromegas technology were also carried out~\cite{bib3}. Recently, additional efforts are pushing the development of resistive strip micromegas detectors in order to increase its robustness in high particle flux environments by limiting spark discharges in the same way as it was done for RPCs. In particular, the MAMMA\footnote{Muon ATLAS MicroMegas Activity} collaboration is developing large area micromegas detectors and introduced the resistive coating technique~\cite{bib8} for the upgrade of the HL-LHC\footnote{High Luminosity Large Hadron Collider (luminosity will be increased by at least a factor 5 reaching up to $L = 5 \cdot 10^{34}$cm$^{-2}$s$^{-1}$)}.

\vspace{0.2cm}

The MAMMA collaboration has investigated new detector prototypes, with different resistive coating topologies. These studies have increased the robustness and stability of micromegas detectors in the presence of an intense and highly ionizing environment, limiting the negative effect of sparks could have on them~\cite{bib5}. Given that the final intention of MAMMA is to participate in the upgrade of ATLAS Muon chambers these prototypes should be proved to be long term radiation resistant. Several aging studies took place in the recent years concerning MPGD and gaseous detectors~\cite{bib4}. However, the introduction of a new technology made of new materials adds the uncertainty of operation during long periods of time in intense particle flux enviroments. The results that we report here concern the first aging tests with this type of detectors using X-ray ionizing radiation. Moreover, it will be completed with future neutron irradiation tests.



\section{Detector prototypes, set-up and characterization}

\begin{figure}[!ht]\begin{center}
\begin{tabular}{ccc}
\includegraphics[width=0.54\textwidth]{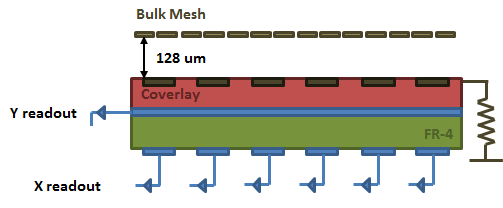} &	&
\includegraphics[width=0.28\textwidth]{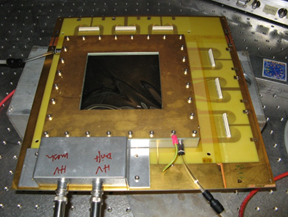} \\
\end{tabular}
\caption{\fontfamily{ptm}\selectfont{\normalsize{ On the left, transversal schematic section of the detector where amplification gap, resistive and read-out strips are observed. On the right a photograph of the detector chamber.  }}}
\label{prototype}
\end{center}\end{figure}

Two identical resistive micromegas detectors, R17a and R17b, built at the PCB CERN workshop [R. de Oliveira] are used in the aging tests taking place at CEA Saclay. These detectors have X and Y copper strips. The top Y-strips have been covered by an insulating \emph{coverlay} layer 60\,$\mu$m thick. The resistive strips 35\,$\mu$m thick are placed above this layer (see Fig.~\ref{prototype}).

Both, resistive and copper strips, have a pitch of 250\,$\mu$m and a width of 150\,$\mu$m. The resistivity along the strips and boundary resistance value was measured during the fabrication process; the first detector, \emph{R17a}, showed a linear resistivity of 45-50\,M$\Omega$/cm, and a boundary resistance of 80-140M$\Omega$. The resistivities obtained for the second detector, $R17b$, were comparable with a linear resistivity of 35-40\,M$\Omega$/cm and a boundary resistance of 60-100\,M$\Omega$.

\begin{figure}[!ht]\begin{center}
\includegraphics[width=0.75\textwidth]{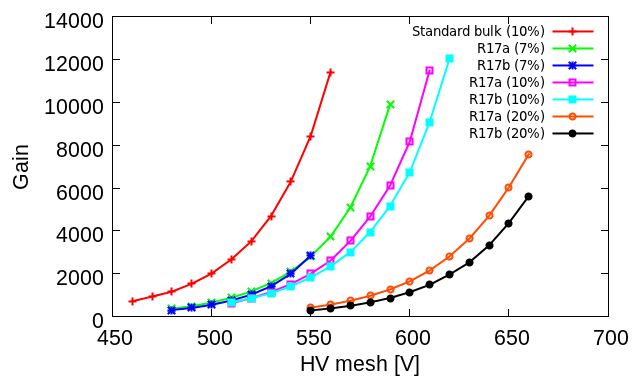} 
\caption{\fontfamily{ptm}\selectfont{\normalsize{ Gain as a function of the mesh amplification voltage for different Argon + CO$_2$ gas mixtures, and comparison with a standard micromegas detector.    }}}
\label{gainCurves}
\end{center}\end{figure}

\vspace{0.2cm}
A first detector characterization in Ar+CO$_2$ mixtures showed the good behavior of the detectors, and the compatibility of the results obtained with both detectors, R17a and R17b. Figure~\ref{gainCurves} shows the gain curves in these different mixtures (amount in volume of CO$_2$ into argon) compared with one obtained with a standard bulk micromegas~\cite{bib9}.

\vspace{-0.8cm}
\section{X-ray generator set-up and characterization}

\begin{figure}[!ht]\begin{center}
\begin{tabular}{ccc}
\includegraphics[width=0.4\textwidth]{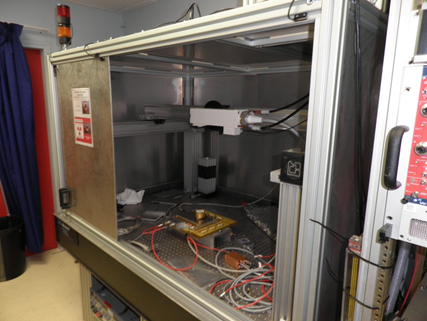} &	&
\includegraphics[width=0.4\textwidth]{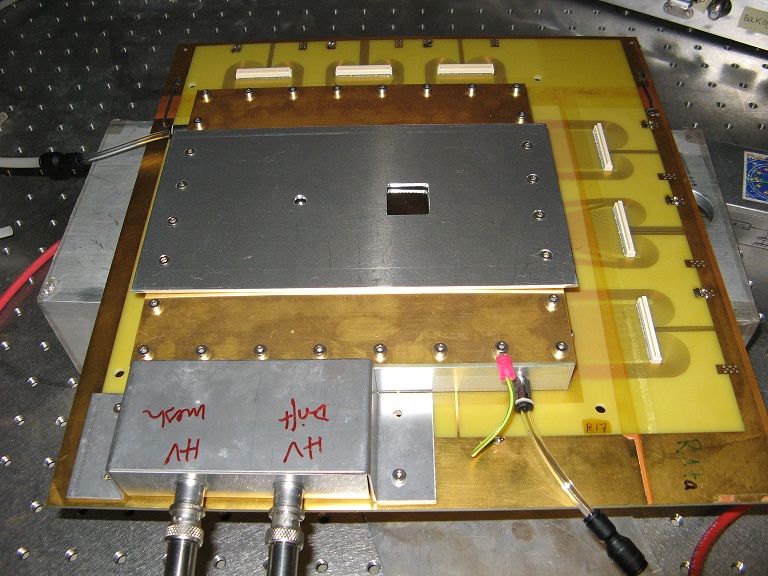} \\
\end{tabular}
\caption{\fontfamily{ptm}\selectfont{\normalsize{ On the left, the X-ray generator cage showing the X-ray generator lamp and the prototype under test placed at the bottom just below. On the right, the detector prototype with the metallic mask on top where the 4\,cm$^2$ squared aperture is visible together with a small hole which was closed during the irradiation tests and that could be used for calibrations at a non-exposed region. }}}
\label{XraySetup}
\end{center}\end{figure}

The prototype under test was placed inside a high intensity X-ray generator cage (see Fig.~\ref{XraySetup}). The X-ray generator consists of an electron gun with an accelerating power of the order of tens of kV and electron currents up to 20\,mA. The electron gun is pointing to a metallic cathode which is excited and emits X-rays, the energy of which depends on the cathode material (8\,keV for Cu).

\vspace{0.2cm}

In order to study the response of the detector for a long exposure time at a specific region of the detector, a mask was prepared by using a aluminum plate with a hole of 4\,cm$^2$ aperture. The plate was placed on top of the detector window, allowing it to be fixed always at the same position (see~Fig.~\ref{XraySetup}). This metallic irradiation mask can be placed in \emph{two} different positions by rotating it 180 degrees.

\vspace{0.2cm}
The X-ray machine was characterized in situ in order to determine the irradiation properties the detector would be subjected to, and to decide the final X-ray generator settings to be used (which allow to tune the illumination rate and spectrum). The X-ray intensity would be too high to determine the X-ray generator energy spectrum and illumination rate with the detector prototype itself. Two methods were used to reduce the intensity: attenuation with a copper foil 175\,$\mu$m thick and use a small collimator of 0.5\,mm diameter.

\vspace{0.2cm}
By using the first method, introducing a copper tape 175\,$\mu$m thick, it was possible to reduce the total rate at the 4\,cm$^2$ area exposed region to few kHz. The attenuation of X-rays by copper and the particular transparency at 8\,keV allows us to characterize the relative rate intensity for the electron current range available. Moreover, one can obtain an estimation of the absolute rate at the detector level by deconvoluting the attenuation by the copper foil at this energy.

\begin{figure}[!ht]\begin{center}
\begin{tabular}{ccc}
\includegraphics[width=0.46\textwidth]{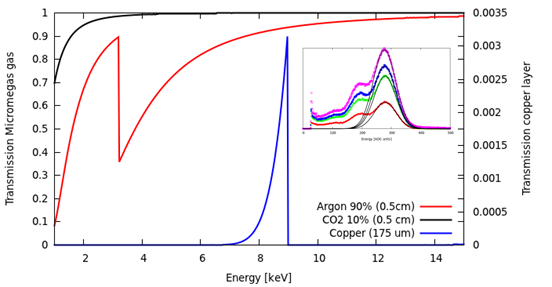} &	&
\includegraphics[width=0.48\textwidth]{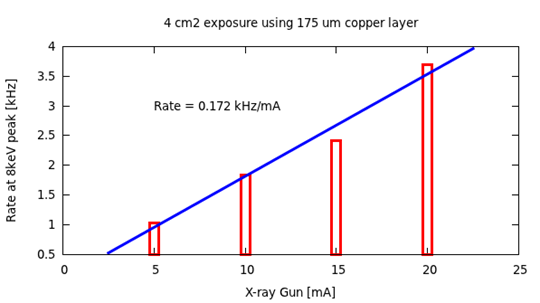} \\
\end{tabular}
\caption{\fontfamily{ptm}\selectfont{\normalsize{ On the left, gas and copper transmission curve (from ref.~\cite{NIST}) together with the measured 8\,keV peaks for different gun electron currents. On the right, the measured rate at different X-ray generator electron currents and the corresponding linear fit.   }}}
\label{copperAtt}
\end{center}\end{figure}

\vspace{0.2cm}
The attenuated rate from copper was determined by measuring the counts inside the resulting 8\,keV peak. The rate was measured for few X-ray generator currents obtaining a expected linear dependency with the electron current of 172\,Hz/mA. Figure~\ref{copperAtt} shows the results obtained, the attenuated rate value as a function of the generator current, together with the 8\,keV spectra obtained and, the gas mixture absorption and copper attenuation.

The copper and gas attenuation length obtained from the NIST database~\cite{NIST}, are integrated in order to determine the relation between the original rate and the attenuated rate measured at the 8\,keV peak. An overall attenuation of a factor of 10$^{-4}$ with respect to the measured rate at 8\,keV was obtained. Fixing the X-ray generator settings at 10\,kV and 10\,mA, the rate at the detector level was about $\sim 40$\,kHz\,mm$^{-2}$ at 8\,keV.

\vspace{0.2cm}

\begin{figure}[!ht]\begin{center}
\includegraphics[width=0.6\textwidth]{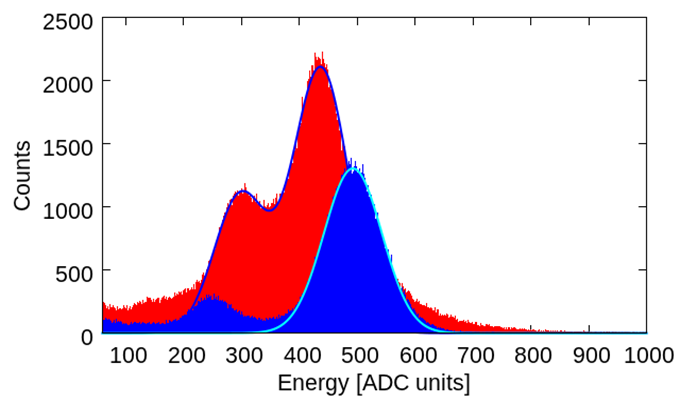} 
\caption{\fontfamily{ptm}\selectfont{\normalsize{Collimated X-ray generator spectrum (red) together with an $^{55}$Fe spectrum (blue) which fixes the ADC conversion to energy in 82.7 ADCs/keV.  }}}
\label{XrayCharacterization}
\end{center}\end{figure}

The second rate measurement was taken using a thin collimator of 0.5\,mm diameter and at a reduced electron gun current of 0.7\,mA at 11.2\,kV. In this measurement it was possible to determine the X-ray flux at the mean cathode energies and the original X-ray generator spectrum, since the reduced overall rate allowed us to minimize the effect of event pile-up. Figure~\ref{XrayCharacterization} shows the spectrum acquired for the X-ray generator compared to an $^{55}$Fe source calibration at the same conditions.

\vspace{-0.2cm}
The X-ray spectrum shows \emph{two} main peaks at 3.5\,keV and 5.3\,keV, which should be mainly produced by the cathode composition. The peak at 5.3\,keV could be related to K-alpha decays coming from chrome and/or vanadium, and the peak at 3.5\,keV could be coming from K-alphas from calcium and potassium that might be associated to a possible contamination of the cathode or the surroundings. The measured collimated X-ray rate of these peaks is 771\,Hz for 3.8\,keV and 1571\,Hz, for 5.3\,keV. Normalizing the rate to an X-ray generator operation current of 10\,mA we obtain the nominal \emph{conversion} rates at these energies to be 5.5\,MHz/cm$^2$ and 11.2\,MHz/cm$^2$.

\section{ Long and intense X-ray aging exposure.} 

The radiation exposure tests aim to accumulate an amount of charge comparable to the values that will be obtained during the lifetime of the HL-LHC. The estimation of the total charge $\sigma_{sLHC}$, produced in the HL-LHC at the muon chambers of ATLAS, is based on the energy deposit $E_{MIP}$ of a MIP (Minimum Ionizing Particle) in 0.5\,cm micromegas conversion gap, in our gas mixture $E_{M.I.P.} \sim 1.25$\,keV, and the detector gain $G$ at the amplification region. The collected charge produced by a MIP is given by 

\begin{equation}
Q_{M.I.P.} =  \frac{E_{M.I.P.}}{W_i} \cdot q_{e^-} \cdot G
\end{equation}

\vspace{0.2cm}

\noindent where $q_{e^-}$ is the electron charge, $W_i = 26.7$\,eV is the mean ionization energy for the gas mixture used. Taking into account a nominal operation gain of 5000, the charge produced per MIP becomes $Q_{M.I.P.} = 37.4$\,fC.

\vspace{0.2cm}

Assuming the expected rate at the HL-LHC future muon chambers will be 10\,kHz/cm$^2$ and taking 5\,years of operation time (200\,days/year), the total generated charge will be $\sigma_{sLHC} = 32.3$\,mC/cm$^2$. The high intensity flux produced at the X-ray generator will allow to produce an amount of charge well above this value for an exposure of few days.

\vspace{0.2cm}
Two different aging measurements took place using \emph{two} different regions of the detector. The gas mixture Ar + 10\% CO$_2$ was used with an equivalent flux of \emph{one} renewal per hour, as it is expected to be at the operation of the HL-LHC chambers. The nominal high voltage conditions remained the same in these \emph{two} periods. The detector gain was set to 5000 by setting the mesh voltage at 540\,V, while the drift field was set to 500\,V/cm.
\subsection{First X-ray aging period.}

The first aging period took place in July 2011. The 4\,cm$^2$ aperture was positioned closer to the APV connectors as it was shown in figure~\ref{prototype}. The X-ray generator was set at a current of 10\,mA and an energy of 11\,keV; these values resulted in an initial mesh current measured with the HV power supply above 500\,nA.

\vspace{0.2cm}

\begin{figure}[!ht]\begin{center}
\includegraphics[width=0.8\textwidth]{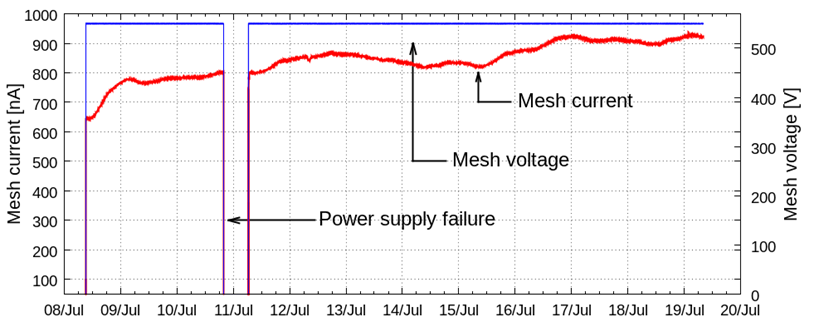} \\
\caption{\fontfamily{ptm}\selectfont{\normalsize{ Mesh current evolution (red) during the first aging period, and mesh voltage (blue) stability during the tests, right scale. }}}
\label{FirstAgeing}
\end{center}\end{figure}

Figure~\ref{FirstAgeing} shows the mesh current evolution at these settings during a period of almost \emph{two} weeks. The mesh current was integrated during the exposure period of 11.2 days resulting in a total charge of 765\,mC in the 4\,cm$^2$ exposed area, thus $\sigma_{aging} = 191.25$\,mC/cm$^2$, and obtaining $\sigma_{aging} \gg \sigma_{sLHC}$. A power supply failure produced a high voltage shutdown during some hours in which the X-ray beam was still operating, not affecting the trend once the field was recovered after a few hours.

During the aging period the mesh high voltage was stable prooving the absence of sparks. The evolution of the mesh current showed an up trend during the overall exposure time leveling at around 900\,nA. The increase from 700 to 900\,nA on the mesh current during the full period is not well understood, and a change too big to be attributed to environmental temperature or pressure changes, and/or beam stability. In fact, the second aging period which will be shown in the next section was motivated by this effect.

\begin{figure}[!ht]\begin{center}
\includegraphics[width=0.7\textwidth]{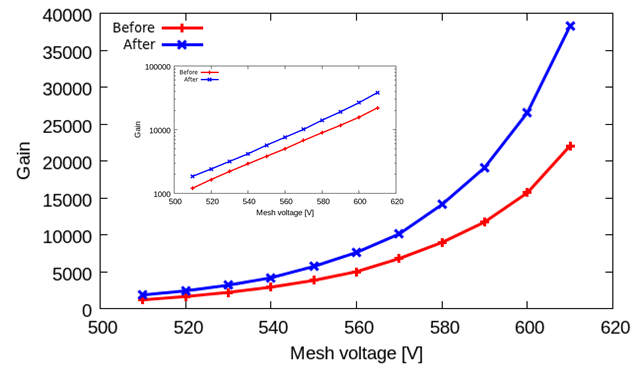} \\
\caption{\fontfamily{ptm}\selectfont{\normalsize{ Gain curves taken with the exposed detector prototype (R17a), before and after the first irradiation period, inset gain in log-scale.  }}}
\label{AgeingGainCurves}
\end{center}\end{figure}

\vspace{0.2cm}
The continuous mesh current during the irradiation proofs the proper operation of the detector. The gain curve, which was measured before and after irradiation (see Fig.~\ref{AgeingGainCurves}), shows the relation of gain versus the amplification field to have the expected behavior, where the increase in gain as a function of mesh voltage is in principle consistent with the change in current during the irradiation period. In any case, the detector gain showed no degradation at the exposed region after the intense X-ray exposure.

\vspace{-0.5cm}
\subsection{Second X-ray aging period.}

\begin{figure}[!ht]\begin{center}
\begin{tabular}{ccccc}
\includegraphics[width=0.27\textwidth]{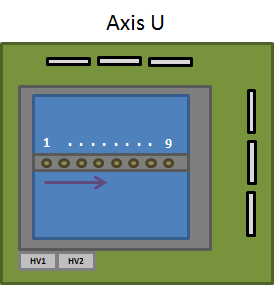} &	&	&	&
\includegraphics[width=0.34\textwidth]{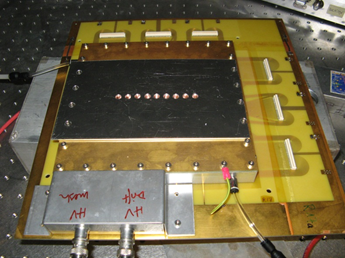} \\
\end{tabular}
\caption{\fontfamily{ptm}\selectfont{\normalsize{ On the left, detector top view schematic with the holes reference mask. On the right, a picture of the mask used to do position calibrations.  }}}
\label{holesSchematic}
\end{center}\end{figure}

In the second period (Oct 2011) a different detector region was exposed, which was accessible by rotating the metallic mask by 180 degrees. While in the first aging period the detector was let alone on the X-ray chamber, in the second aging period the second prototype (R17b) was placed inside the X-ray generator chamber protected from radiation, and sharing the same gas flow in parallel. In the first period gain curves were taken just before and after the irradiation, while at the second period reference gain values were taken with the R17b detector at the same conditions during the irradiation measurement. Moreover, in order to study if there was any relative effect on gain at the exposed region a metallic mask with several equi-distant holes was prepared~(see Fig.~\ref{holesSchematic}). 

\vspace{0.2cm}

\begin{figure}[!hb]\begin{center}
\includegraphics[width=0.92\textwidth]{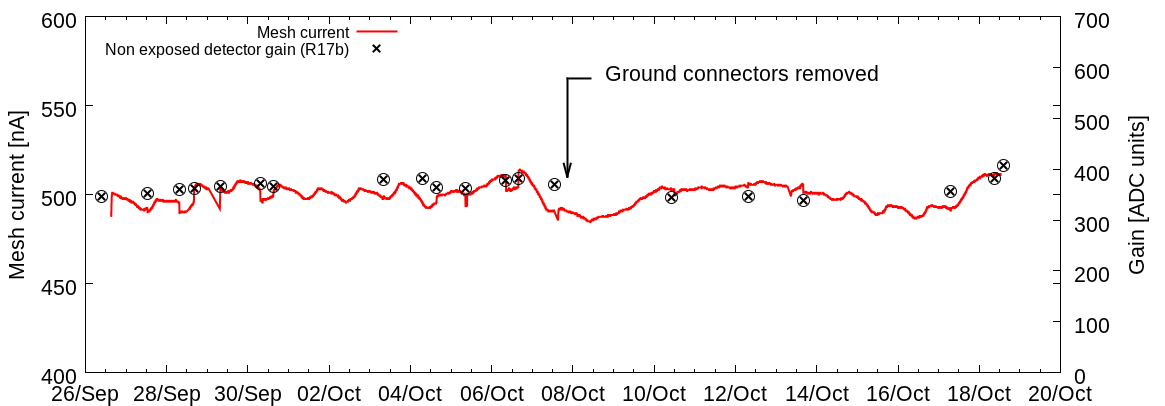} \\
\caption{\fontfamily{ptm}\selectfont{\normalsize{ Mesh current evolution provided by the high voltage power supply (red line) and the R17b gain control measurements with R17b detector (black circles). }}}
\label{SecondAgeing}
\end{center}\end{figure}

It was thought that the mesh current effect observed in the first aging test could be related to the absence of grounding connectors at the standard strip ends (even-though the top-plane with resistive strips is already grounded). Grounding connectors were prepared to directly drive the strips to ground. The current evolution at the second aging (see Fig.~\ref{SecondAgeing}) remains at a constant level of 500\,nA where the expected daily gain fluctuations are observed\footnote{ The current value was preset to 500\,nA by re-adjusting the X-ray generator settings}. After several days, the connectors were removed to observe the effect on the mesh current showing a kind of integrating effect of the daily fluctuations but not in the general current trend which remained stable. The different gain measurements at the auxiliary detector R17b during the second aging period corroborates the detector gain stability.

During this second aging period the integrated mesh current is 918\,mC in 4\,cm$^2$, produced over 21.3 effective days. Three gain measurements of both detectors took place first before starting the tests, second in the middle of the aging period when the connectors were changed and third after the exposure. In order to study the relative gain homogeneities before and after exposure, the gain is normalized to the maximum gain at each set of measurements.

\begin{figure}[!ht]\begin{center}
\begin{tabular}{ccc}
\includegraphics[width=0.45\textwidth]{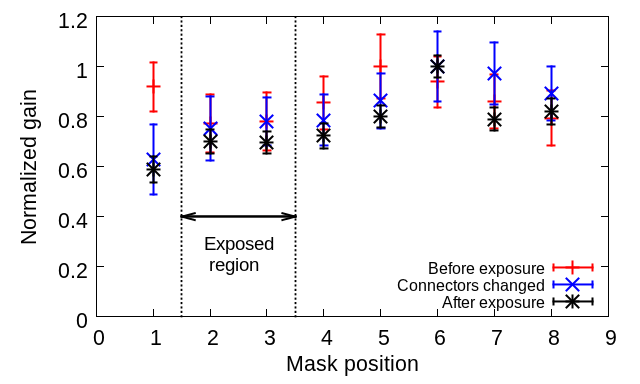} & &
\includegraphics[width=0.45\textwidth]{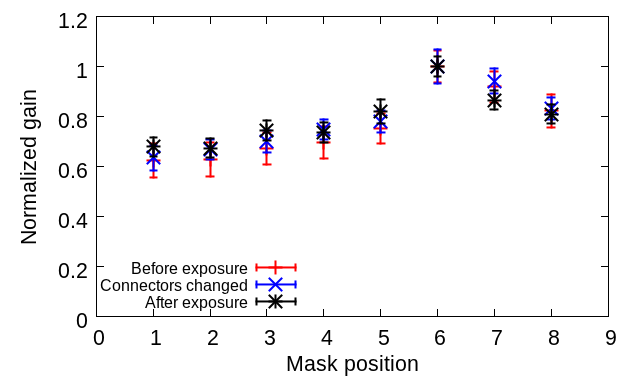} \\
\end{tabular}
\caption{\fontfamily{ptm}\selectfont{\normalsize{ Gain measurements as a function of position before the second irradiation period, during, and after the irradiation period. The exposed detector R17a (left) and non-exposed detector R17b (right). The exposed region is indicated in the left plot. }}}
\label{positionGain}
\end{center}\end{figure}

Figure~\ref{positionGain} shows the relative gain at each hole position for these \emph{three} set of measurements. This result shows that the gain profile in both detectors has the same structure, and is compatible with previous measurements, at different aging periods. Moreover, the exposed detector region do not show a relevant difference in relative gain with respect to the non-exposed regions, a definitive proof that there was no effect on the original behavior of the detector.

\section{Conclusions and future work}

We have exposed a micromegas prototype based on the resistive strip technology to an intense X-ray beam observing no deterioration of the detector at the aging regions. The two exposed regions were under irradiation long enough to accumulate an operating charge that is far above the levels that will be reached at the future HL-LHC during 5 years of operation.

\vspace{0.2cm}

Additional aging tests with an intense beam of cold neutrons is taking place in the Orph\'ee reactor at CEA Saclay. These measurements will allow us to verify the capability of this new detector technology to stand this kind of irradiation and thus if there is any effect on the detector materials owing to nuclear interactions.


\end{document}